\documentclass[pra,twocolumn,showpacs,superscriptaddress,floatfix,amssymb]{revtex4}
\usepackage{graphicx}
\usepackage{wasysym}

\newcommand{\ket}[1]{\ensuremath{\left|{#1}\right\rangle}}
\newcommand{\bra}[1]{\ensuremath{\left\langle{#1}\right|}}
\newcommand{\braket}[1]{\ensuremath{\left\langle{#1}\right\rangle}}
\newcommand{\neigh}[1]{\ensuremath{{\cal N}(#1)}}

\begin{document}

\title{Mapping the spatial distribution of entanglement in optical lattices}

\author{Emilio Alba}
\affiliation{Instituto de F{\'\i}sica Fundamental, CSIC, Calle Serrano 113b, Madrid E-28006, Spain}
\author{G\'eza T\'oth}
\affiliation{Department of Theoretical Physics, The University of the Basque Country, P.O. Box 644, E-48080 Bilbao, Spain}
\affiliation{IKERBASQUE, Basque Foundation for Science, E-48011 Bilbao, Spain}
\affiliation{Research Institute for Solid State Physics and Optics, Hungarian Academy of Sciences, P.O. Box 49, H-1525 Budapest, Hungary}
\author{Juan Jos\'e Garc{\'\i}a-Ripoll}
\affiliation{Instituto de F{\'\i}sica Fundamental, CSIC, Calle Serrano 113b, Madrid E-28006, Spain}
\email{jjgr@iff.csic.es}

\begin{abstract}
In this work we study the entangled states that can be created in bipartite two-dimensional optical lattices loaded with ultracold atoms. We show that using only two sets of measurements it is possible to compute a set of entanglement witness operators distributed over arbitrarily large regions of the lattice, and use these witnesses to produce two-dimensional plots of the entanglement content of these states. We also discuss the influence of noise on the states and on the witnesses, as well as the relation to ongoing experiments.
\end{abstract}


\maketitle

\section{Introduction}
The quantum engineering of useful many-body states and the characterization of their entanglement properties are two of the most challenging topics in Quantum Information Science, both theoretically and experimentally. In the laboratory, the creation of entangled states has been addressed in two ways. The first one 
starts from the control of individual quantum systems, let it be photons~\cite{pan00,bourennane04,walther05,lu07}, neutral atoms~\cite{widera04,anderlini07} or ions~\cite{haffner05,leibfried05}, and aims at the creation of larger and larger many-body states. The second one consists of taking large numbers of these components (e.g., $10^3-10^6$ atoms) and studying collective degrees of freedom~\cite{julsgaard01}. It thus seems that one has to make a compromise between having large entangled states or having a fine-grane knowledge of the properties of the state.

In this work we show that there is an intermediate approach, by which it is possible to gain local information about a very large entangled state. More precisely, we introduce a family of operators that allow for obtaining lower bounds on the fidelity or detecting multipartite entanglement in regions of a two-dimensional graph state. The entanglement witnesses \cite{Horodecki1996,Terhal2000,Lewenstein2000,Cirac2002,Guhne2002}
are optimized for setups with ultracold atoms in 2D bipartite lattices, in which one now has access to the state of individual atoms~\cite{bakr09,soderberg09}. Remarkably, our witnesses only require the simultaneous measurement of all atoms, but with a postprocessing of the measurement statistics it provides a map of the quality and multipartite entanglement of the many-body state.

\begin{figure}[t]
  \centering
  \includegraphics[width=\linewidth]{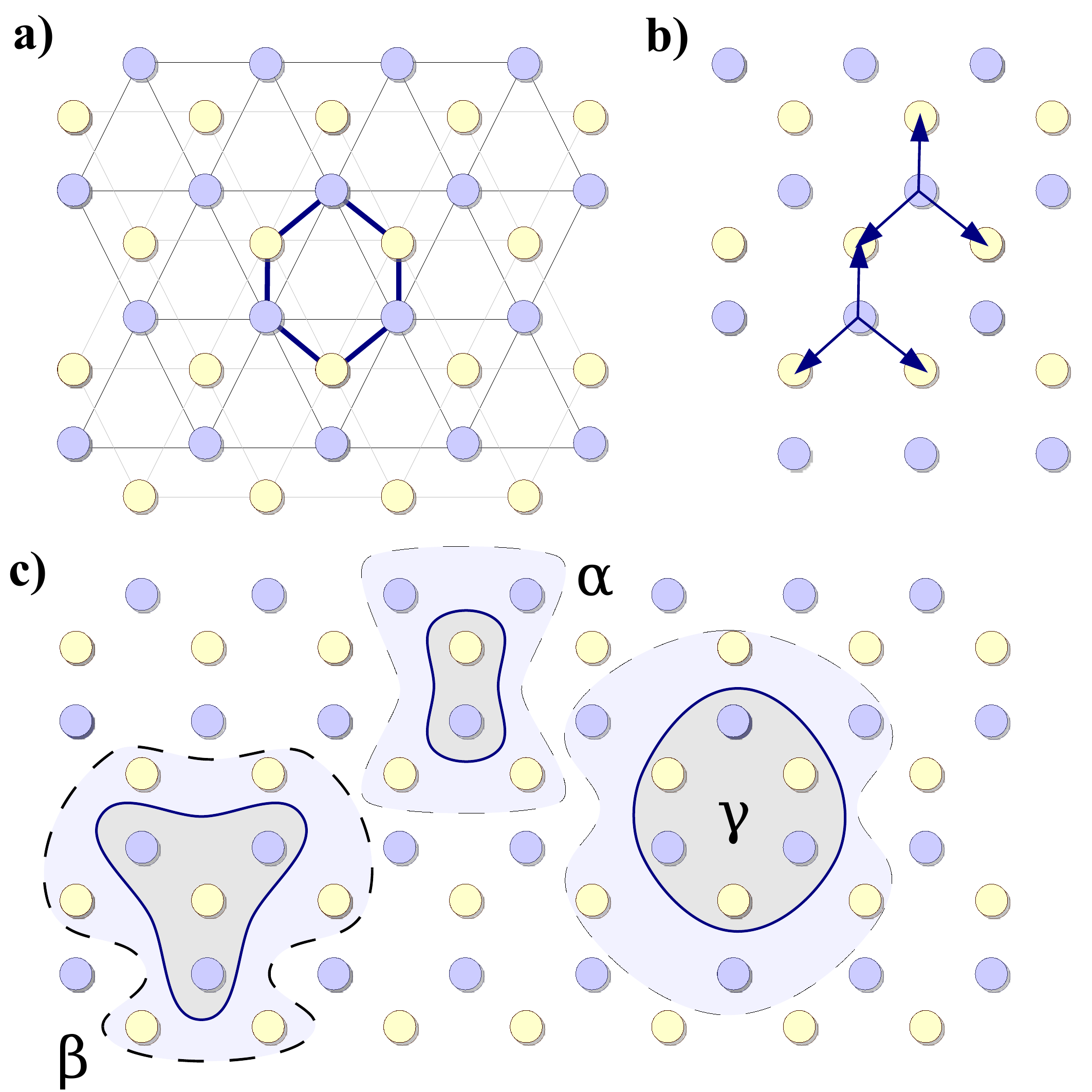}
  \caption{(Color online) (a) We work with two species of atoms, Cs and Li, trapped in two independent triangular sublattices, which together form a honeycomb lattice. (b) A graph state can be generated moving one of the sublattices along three different directions. 
   (c) In this work we analyze the properties of localizable multipartite entanglement in small sets of 2 $(\alpha),$ 4 $(\beta),$ 6 $(\gamma),$ or more spins. Each of these regions, $\Omega,$ is connected with its boundary, $\partial\Omega,$ by two-qubit unitaries.}
  \label{fig:setup}
\end{figure}

This paper is structured as follows: in Sec. \ref{section2} we review the experimental techniques available to create graph states in optical lattices.  We present observables that characterize the state and act as entanglement witnesses in section \ref{section3}, in which we also face the main difficulties associated to this method. We find that even under decoherence states useful for quantum computation can be found, and analyse simple observables that bound the fidelity of the state. Finally, in Sec. \ref{section4}, we perform numerical simulations of a cluster state subject to different noise sources, demonstrating that the entanglement witness is capable of detecting those errors.

\section{Experimental generation of stabilizer states}\label{section2}

The original method for creating graph states with neutral atoms \cite{Briegel01} was based on filling state-dependent lattices with one atomic species and using controlled collisions~\cite{mandel03,Jaksch99}. In contrast, we will develop our ideas building on the experimental setup from Ref.~\cite{soderberg09,klinger10}, which traps two different species of atoms in two coexisting optical lattices, one of which can be moved. This setup combines a diffraction mask with a powerful microscope objective, which projects two similar triangular lattice patterns on its focal plane. Using two light beams with different frequencies, the experiment may trap lithum and cesium atoms in two independent lattices that can be moved at will along the plane that confines the atoms. As shown in Fig.~\ref{fig:setup}a, we can contemplate the Cs and Li arrangements as the triangular sublattices of a larger honeycomb lattice where each Cs atom is surrounded by three Li atoms, and vice versa; and each atom acts as one qubit. Since our lattice is bipartite by construction entanglement can be created using a small number of steps, equal to the coordination number of the full lattice. Continuing with this example, one has to move one sublattice three times so that each Cs atom approaches each of its neighboring Lithium atoms~[Fig.~\ref{fig:setup}b], suffering a controlled collision~\cite{mandel03} or an engineered interaction~\cite{soderberg09}. A fundamental difference with previous setups~\cite{Briegel01,mandel03,Jaksch99} is that the sublattice now moves as a whole, regardless of the internal states of the atoms. If the lattices are very deep and the atom-atom interaction is strong enough, this can be done with great precision.

To fix ideas we will assume that the entangling operation between atoms in different sublattices is a CZ gate, $U_{CZ} = \exp\left(-i\frac{\pi}{4}\sigma^z_{\mathrm{Cs}}\sigma^z_{\mathrm{Li}}\right).$ After three parallel sets of operations, beginning with a product state, $(\ket{0}+\ket{1})^{\otimes N},$ we will arrive at a graph state
\begin{equation}
  \ket{G_{\hexagon}} \sim \prod_{i\in A} \prod_{j\in \neigh{i}} U_{CZ}^{(i,j)} \left(\ket{0}+\ket{1}\right)^{\otimes N_A+N_B},
\end{equation}
where $A$ and $B$ denote the Cs and Li sublattices and $\neigh{i}\subset B$ is the set of nearest neighbors to the potential well $i.$ Note that if instead of using the control-phase one implements a control-NOT, $U_{\mathrm{CNOT}} = \left(1+\sigma^z_{\mathrm{Li}}\right) - \left(1-\sigma^z_{\mathrm{Li}}\right)\sigma^x_{\mathrm{Cs}},$ where the Cs absorbs the parity of its neighbors, we obtain what we call a ``parity'' multipartite entangled state
\begin{equation}
  \ket{P_{\hexagon}} \sim \prod_{i\in A} \prod_{j\in \neigh{i}} U_{\rm CNOT}^{(i,j)} \left(\ket{0}+\ket{1}\right)^{\otimes N_A+N_B}.
\end{equation}

\section{Entanglement  witnesses}\label{section3}
\subsection{Global fidelity of graph states}

All the states that we can create using the previous operations belong to the family of stabilizer states. In both cases we have a complete set of $N_A+N_B$ local observables, the stabilizing operators, $g_i,$ that may take values $\{-1,+1\},$ and for which the states $G_{\hexagon}$ and $P_{\hexagon}$ are eigenstates with eigenvalue $+1$ on all sites~\footnote{The $g_i$ operators are the generators of the so called stabilizer group~\cite{gottesman98}.}. For instance, in the case of the graph state we have
\begin{equation}
  g_i \ket{G_{\hexagon}}= +1 \ket{G_{\hexagon}},\quad\forall i \in A \cup B.
\end{equation}
with the stabilizing operators $g_i = \sigma^x_i\prod_{j\in\neigh{i}} \sigma^z_j.$ In general, given a set of lattice sites, $\Omega,$ we can construct a projector onto a stabilizer state containing those sites
\begin{equation}
  P_\Omega = \prod_{i\in \Omega} \frac{1}{2}(1 + g_i).\label{P-function}
\end{equation}
In theory we can use this projector to compute the fidelity of our experimentally realized state, $\rho,$ which is probably mixed, with respect to the objective $G_{\hexagon}$ or $P_{\hexagon},$
\begin{equation}
  F_{A\cup B} = \mathrm{tr}(P_{A\cup B} \rho),
\end{equation}
where the region under study now encloses the $A$ and $B$ sublattices.  However in practice this is already impossible for a few qubits, since the evaluation of $F_{\Omega}$ requires us to measure $2^{N_A+N_B}$ different observables coming from all possible products of the $g_i$ operators. The difficulty of this task seems to be tantamount to performing a full tomography of the mixed state $\rho.$

\subsection{Localizable fidelity}
Instead of following this very complicated route, we will focus on two simpler questions, which are intimately related: (i) a notion of local fidelity to the stabilizer state and (ii) the existence and detection of genuine multipartite entanglement \cite{Sackett2000,Acin2001}
in the lattice. In both cases we can extract a number, let it be a fidelity or the expectation value of an entanglement witness, $F(i)$ or $W(i),$ which is distributed over the 2D lattice of sites. With those numbers we can study the distribution of entanglement and how much our state has been affected by noise or decoherence.

Our notion of ``localizable fidelity'' builds on the fact that given a simply connected set of sites, $\Omega,$ and a perfect graph state, $\ket{G_{\hexagon}},$ we can extract another perfect graph state in that region. One way to achieve this is measuring the boundary qubits, $\partial \Omega,$ [See Fig.~\ref{fig:setup}c] and, depending on the outcome of those measurements, performing phase gates on the qubits that were immediately connected to them. An alternative but completely equivalent way is to disentangle the boundary with the same two-qubit unitaries we used to build the state
\begin{equation}
  \rho_\Omega = \mathrm{tr}\left(\prod_{i\in\partial\Omega}
    \prod_{j\in\neigh{i}} U_{CZ}^{(i,j)} \rho_{A\cup B}\right).
\end{equation}
The most important idea is that this procedure still can be applied if the initial state of the atomic ensemble is mixed, $\rho_{A\cup{B}},$ due to decoherence. In this case the fidelity of the final state is related to the same observable that we found before, that is
\begin{equation}
  \label{eq:6}
  F_\Omega = \bra{G_\Omega}\rho_\Omega\ket{G_\Omega}
  = \mathrm{tr}\left(P_\Omega\rho_{A\cup{B}}\right),
\end{equation}
the fidelity of the final state only depends on how close $\rho_{A\cup{B}}$ is to the eigenstates of the stabilizing operators that cover the region \textbf{and} the boundary, $\Omega\cup{\partial\Omega}.$ The final observation is that the fidelity $F_\Omega$ gives us not only local information about how close our state is to the graph state, but it is also a witness for genuine multipartite entanglement in that region, $W_\Omega = \frac{1}{2}\openone - P_\Omega$ \cite{tothprl05}.

\begin{figure*}[t!]
  \includegraphics[width=\linewidth]{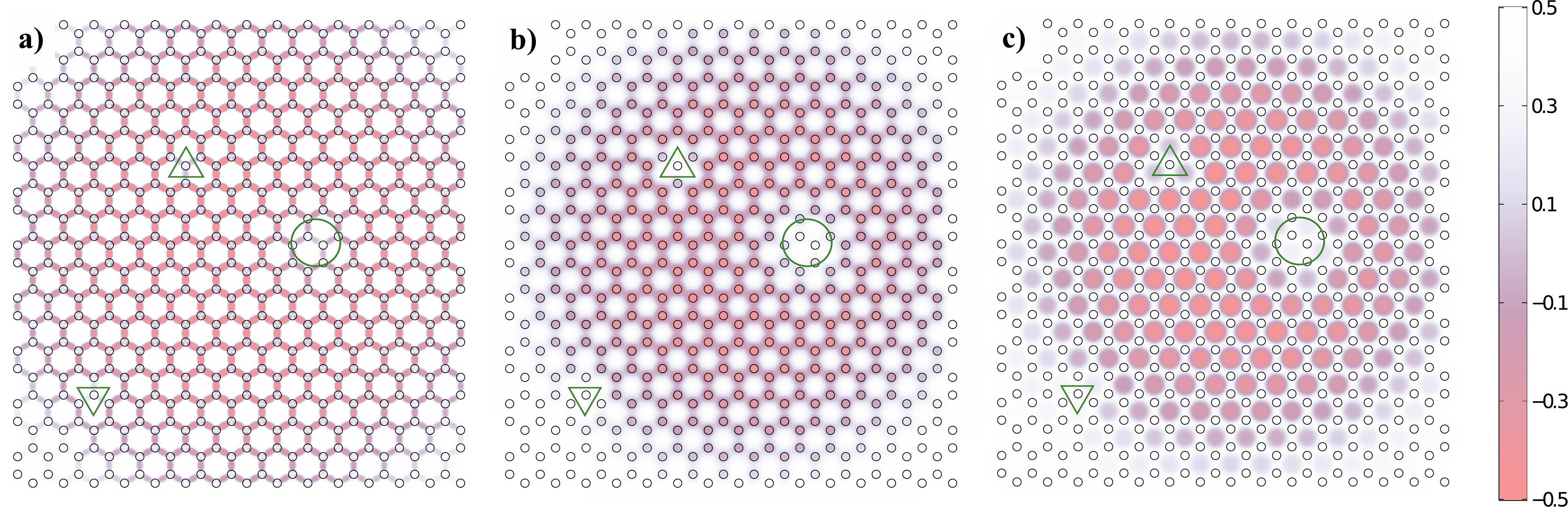}
  \caption{(Color online) Two-dimensional distribution of the entanglement witnesses for (a) two, (b) four and (c) six particle arrangements, $\alpha,\beta$ and $\gamma$ from Fig.~\ref{fig:setup}c, respectively. The value of the witness is color coded on the (a) links, (b) atoms or (c) center of the plaquette. A a negative value of the witness (red) denotes existence of bi- or multi-partite entanglement. All pictures present the same defects, consisting on two empty sites (triangles), atoms subject to strong dephasing (circle), and an increase of phase gate errors towards the edges of the trap.}
  \label{fig:2dplots}
\end{figure*}

\subsection{Optimized witnesses}
However, even if $\braket{W}<0$ detects entanglement, the evaluation of this quantity seems to require a number of measurements that increases exponentially with the number of qubits. We thus need another ingredient, which is obtained by writing the fidelity as a product of two operators constructed from stabilizing operators corresponding to different sublattices, $P_\Omega = P_{\Omega\cap{A}}P_{\Omega\cap{B}},$ and introducing a new operator
\cite{toth05}
\begin{equation}
  \tilde{P}_\Omega = P_{\Omega\cap{A}}+P_{\Omega\cap{B}}-\openone
  \leq P_\Omega,
\end{equation}
This equation can be readily verified in the basis that diagonalizes both $P_{\Omega\cap{A}}$ and $P_{\Omega\cap{B}}$, where the eigenvalues of the projectors can only be $0$ or $+1$.

This observable provides a lower bound for the fidelity
\begin{equation}
  F_\Omega \geq \langle{\tilde{P}}_\Omega\rangle,
\end{equation}
and can be used to construct an entanglement witness
\begin{equation}
  \tilde{W}_\Omega = \frac{1}{2}\openone - \tilde{P}_\Omega.
  \label{WOmega}
\end{equation}
The advantage is that now the quantities $\langle{P_{\Omega\cap{A}}}\rangle$ and $\langle{P_{\Omega\cap{B}}}\rangle$ can be extracted from just two settings of measurements. In particular, for the graph state one such expectation values
\begin{equation}
\langle{P_{\Omega\cap{A}}}\rangle = \left\langle\prod_{i\in \Omega\cap{A}}\frac{1}{2}\left(1 + \sigma^x_i\prod_{j\in N(i)} \sigma^z_j\right)\right\rangle,
\end{equation}
is obtained measuring $\sigma^x$ in all Cs atoms ($i\in\Omega\cap{A}$) and $\sigma^z$ in the Li ($j\in N(i)$), while the other expectation value is obtained with the opposite measurement basis. Note also that postprocessing the \textbf{same} set of measurement results we can compute the values $\langle \tilde{P}_{\Omega}\rangle$ for any region $\Omega,$ which allows us to produce two-dimensional displays of the distribution of localizable fidelity or the multipartite entanglement witness.

\section{Simulations}\label{section4}
This section features a numerical simulation of a realisation of our method in an experiment taking possible practical sources of error into account. We have studied the degradation of the expectation value of the witness $\tilde{W}_\Omega$ given in Eq.~(\ref{WOmega}). In general it is not possible to compute easily the change of $\langle P_{\Omega}\rangle,$ but we will take advantage of the facts that the witness is the sum of two functions of stabilizing operators corresponding to {\it different} sublattices and that for our sources of noise these expectation values have simple expressions, such as $\langle
P_{\Omega\cap{A,B}}\rangle=\prod_{i\in\Omega\cap{A,B}}(1+\langle g_i \rangle)/2,$ in which only the expectation values of isolated stabilizers appear, $\langle g_i\rangle.$ As explained in Appenndix~\ref{Appendix}, we have considered various types of noise~\cite{benenti07} using the quantum channel formalism to compute the changes in $\langle g_i\rangle.$
\begin{itemize}
\item[(i)] Dephasing, which is due to fluctuations in the energy levels of the atoms due to external fields, $\epsilon_j(\rho) = \int d\theta_j \exp(-i\sigma^z\theta_j) \rho \exp(i\sigma^z\theta_j) p(\theta_j).$ This map is repeated on all sites, with site dependent uniformly distributed random phases in $[-\epsilon_j,\epsilon_j],$ degrading the stabilizing operator $\langle g_i\rangle \to \langle g_i \rangle \prod_{i\in\Omega}\sin(2\epsilon_{i})/2\epsilon_i.$
\item[(ii)] Imperfections in the gates that entangle pairs of sites, $U_{CZ}^{(j,k)} \to U_{CZ}^{(j,k)} \exp(i\theta_{jk} \sigma^z_j\sigma^z_k),$ where $\theta_{jk}$ are again random variables, uniformly distributed in $[-\epsilon_{jk},\epsilon_{jk}].$ This error introduces a factor in the expectation value of the stabilizing operators, $\langle g_i\rangle \to \langle g_i \rangle \prod_{ j \in N(i) }\frac{1}{2}(1 + \sin(2\epsilon_{ij})/2\epsilon_{ij}),$ plus other terms that do not contribute to the witness~(\ref{WOmega}).
\item[(iii)] Atom loss, which introduces a new state in the lattice,  the hole $|h\rangle.$ 
In practice, it can be described by $\epsilon_{\mathrm{AL}}(\rho) = (1-p)\rho + p \ket{0}\bra{0}.$
\item[(iv)] Spontaneous emission $\epsilon_{\mathrm{SE}}(\rho) = (1-p)\rho + p \ket{0}\bra{0}.$
\item[(v)] The completely depolarizing channel $\epsilon_{\mathrm{DP}}(\rho) = (1-p)\rho + \frac{p}{2}\openone.$

\end{itemize}
The last three sources of error have the same effect, $\langle g_i\rangle\to (1 - p)\langle g_i\rangle.$

With these types of noise and decoherence we studied the evolution of our witness operators and the overall description of a potential experiment using them. The results are shown in Figs.~\ref{fig:2dplots}a-c, where we plot the values of $\tilde{W}_\alpha, \tilde{W}_\beta$ and $\tilde{W}_\gamma,$ interpolated using smooth functions that are centered and cover the affected regions, $\alpha, \beta$ and $\gamma,$ of two, four and six qubits. The result is a two-dimensional map of the entanglement content, where the value of the witness is color-coded either (a) on the link between two atoms, $\alpha,$ (b) on the central atom and the star surrounding it, $\beta,$ or (c) on the center of the six-atom plaquette, $\gamma.$ In these particular plots we have combined all sources of decoherence, making some of them more relevant in different regions of the lattice.  We have introduced a region of atoms subject to strong dephasing induced by a focused laser, covering the area marked by a circle. We have also emptied two sites, surrounded by a triangle. These empty holes are numerically equivalent to having spontaneous emission with 100\% probability.  Finally, we have assumed that the phase gate is 100\% accurate in the center of the lattice and acquires a 10\% error at the boundary of the lattice.

We already appreciate interesting features in these simple simulations. The first one is that bipartite entanglement is less affected by noise than multiqubit arrangements. While we can reconstruct a Bell state close to the boundary with an 80\% error, the four- and six-qubit states only have an appreciable value of the witness when the CZ gate is above 90\% fidelity. The second feature is that the effect of local errors remains local. The sites, bonds and plaquettes that share one or more qubits with the regions affected by atom loss or strong decoherence (circle and triangle in the plot) have positive values of the witness and do not have significant entanglement. However, one site or plaquette away from the region of influence of those defects, the witnesses recover their large negative value.

\section{Conclusions}
In summary, we have presented a simple scheme for detecting bipartite and multipartite entanglement in two-dimensional lattices with ultracold atoms. The present study admits a straightforward generalization not only to other bipartite lattice setups, such as square lattices, but also to other interaction schemes $(U_{\rm CNOT})$, or to displacing each Cs atom not three, but one or two times. First of all, if the Cs atoms move along two directions, the result is an array of linear cluster states, with an entanglement witness that is a generalization of the previous ones, and that again relies only on two-measurement settings~\cite{tothprl05}. If instead we move each Cs atom only once, then the Cs-Li interact in pairs forming a macroscopic number of disconnected two-qubit singlets. In this case we do not need a witness but can rather compute the expectation value of the projector
\begin{equation}
P=\frac{1}{4}\left(\openone + \sigma^x_{\rm Cs}\sigma^x_{\rm Li}+\sigma^y_{\rm Cs}\sigma^z_{\rm Li}+\sigma^z_{\rm Cs}\sigma^y_{\rm Li}\right),
\end{equation}
using three experimental settings.

We must remark that our scheme only uses the fact that the lattice is bipartite and that it is possible to simultaneously measure the state of all lattice sites in each sublattice independently. In particular, while we have focused on a two-species setup~\cite{soderberg09,klinger10}, exactly the same protocols and measurement schemes can be used using the state-dependent optical lattices in previous experiments~\cite{mandel03}, combined with the new optics that allows imaging individual lattice sites~\cite{greiner09}. The only difference is that since we do not have different atoms on different sublattices, the measurement protocol has to be preceded by a global and local rotation of one sublattice to change its measurement basis. This is not too complicated and can be done using two counterpropagating laser beams in an optical lattice configuration, such that their maxima of intensity concide with just one sublattice.

Our proposal represents the first experimentally realizable scheme for mapping out the entanglement distribution and fidelity of a very large many-body correlated state. It also opens the path for the experimental detection of very large cluster states, a task which so far was not achievable using ultracold atoms in optical lattices, but which becomes possible for ongoing experiments using two species of atoms and holographically generated trapping potentials~\cite{soderberg09}. In particular, we want to remark that the family of graph states in honeycomb lattices is a universal resource for measurement-based quantum computation and that our scheme can be used to isolate regions of high fidelity in such resources.

\section*{Acknowledgements}

G.T thanks the support of the Spanish MICINN (Consolider-Ingenio 2010 project ''QOIT'', project No. FIS2009-12773-C02-02), the Basque Government (project No. IT4720-10), and the ERC Starting Grant GEDENTQOPT. E.A. and J.J.G.R acknowledge support from the Spanish MICINN project No. FIS2009-10061, CSIC Project Ref. 200850I044 the CAM project QUITEMAD. E.A. thanks support from MEC grant FPU2009-1761.

\appendix
\section{Positive maps and Noise sources} \label{Appendix}

Any physical operation on a quantum state must be a trace preserving
positive map, which maps density matrices into density matrices. Furthermore,
such operators admit a unique decomposition using a set of operators\[
\varepsilon(\rho)=\sum_{k}A_{k}\rho A_{k}^{\dagger},\]
with the property $
\varepsilon(\openone)=\openone.$

This description admits a generalization to expectation values. In other words, we also have a positive
map description in the Heisenberg picture, where operators/observables
and not states are changed. Using the definition $
\langle\Theta\rangle=\mathrm{tr}\{\Theta\rho\},$
the change in the expectation value can be expressed as\[
\langle\Theta\rangle_{\varepsilon(\rho)}=\mathrm{tr}\{\Theta\sum_{k}A_{k}\rho A_{k}^{\dagger}\}=\langle\tilde{\varepsilon}(\Theta)\rangle\]
where $
\tilde{\varepsilon}(\Theta)=\sum_{k}A_{k}^{\dagger}\Theta A_{k}.$

We now want to estimate the effect of different positive maps on our entanglement
witnesses. We will first focus on local error sources. It is important to observe that the expected values we want to calculate are of the general form \[
P=f(\sigma_{i\in A}^{\alpha},\sigma_{i\in B}^{\beta}),\quad\alpha,\beta\in\{x,z\},\]
that is, they are functions of the same observables on each sublattice. This
means that under local error sources the following relation applies:\[
P'=f\left(\tilde{\varepsilon}(\sigma_{i\in A}^{\alpha}),\tilde{\varepsilon}(\sigma_{i\in B}^{\beta})\right).\]
Therefore it suffices to compute how the operators change under the different local error sources. In each case, the decoherence channel will change the effective value of the stabilizer expectation value.

\subsection{Dephasing}

In this noise source we have an average over random phases\[
\tilde{\varepsilon}(\sigma^{x})=\int e^{-i\phi\sigma^{z}}\sigma^{x}e^{i\phi\sigma^{z}}p(\phi)d\phi.\]
If the distribution $p(\phi)$ is symmetric, then \[
\tilde{\varepsilon}(\sigma^{x})=\int\left[\cos(2\phi)+i\sin(2\phi)\sigma^{z}\right]\sigma^{x}p(\phi)d\phi=(1-\epsilon_{i})\sigma^{x},\]
with some error factor $\epsilon_{i}.$ Since the $\sigma^{z}$ operators
are not affected, it is legitimate to say that the map induces a global change in the expected value $g_{i}\to(1-\epsilon_{i})g_{i}.$

\subsection{Particle loss}

This positive map has the
form\[
\epsilon(\rho)=(1-p)\rho+p|0\rangle\langle0|,\]
 which we can also write in Kraus form using the operators\[
A_{0} =  (1-p)\openone,
A_{1} = p|0\rangle\langle0|,
A_{2} =  p|0\rangle\langle1|.
\]This means that the operators transform as \[
\tilde{\varepsilon}(\Theta)=(1-p)\Theta+p\langle0|\Theta|0\rangle\openone. \]
Thus the stabilizer operators are modified\[
g_{i}\to(1-p)^{N}\sigma_{i}^{x}\prod_{j\in N(i)}\sigma_{j}^{z}+g_{i}^{\perp},\]
where N is the number of qubits in the stabilizer operator ($4$ in our
case for the honeycomb lattice) and the $g_{i}^{\perp}$ contain terms that 
are going to vanish because they can be written in the form 
$g_{i} \sigma_{i}^{x}\prod_{j\in N'(i)}\sigma_{j}^{z}$ or $g_{i} \prod_{j\in N'(i)}\sigma_{j}^{z}$
with $N'(i) \subseteq N(i),$ so that their expectation values are zero.


\subsection{Errors in the gates}

We can proceed similarly though some subtleties are to be taken into account. First
of all we realize that instead of transforming the state we can transform
the stabilizer operators which appear in the expectation value\[
g_{j}\to e^{-i\sum_{k\in N(j)}\epsilon_{jk}\sigma_{j}^{z}\sigma_{k}^{z}}g_{j}e^{+i\sum_{k\in N(j)}\epsilon_{jk}\sigma_{j}^{z}\sigma_{k}^{z}}.\]
It can be seen that this is equivalent to performing the same transformation
only on the $\sigma_{j}^{x}$ operator \[
\sigma_{j}^{x}\to\prod_{k}\left[\cos(2\epsilon_{jk})+i\sin(2\epsilon_{jk})\sigma_{j}^{z}\sigma_{k}^{z}\right]\sigma_{j}^{x},\]
Note that since we only have $\sigma^{x}$ operators in one sublattice and $\sigma^{z}$ on
the other, the phases that we have here are uncorrelated among different
$\sigma^{x}$ operators. Furthermore, any term that contains a $\sigma^{z}$ operator
vanishes once we take expectation values, which means
that we can replace $
\sigma_{j}^{x}\to(1-\epsilon_{j})\sigma_{j}^{x},$
where $
\epsilon_{j}=\prod_{k}\int\epsilon_{jk}p_{jk}(\epsilon_{jk})d\epsilon_{jk}.$
This shows that the outcome is a global reduction of the stabilizer expectation value.


\begin{thebibliography}{27}
\expandafter\ifx\csname natexlab\endcsname\relax\def\natexlab#1{#1}\fi
\expandafter\ifx\csname bibnamefont\endcsname\relax
  \def\bibnamefont#1{#1}\fi
\expandafter\ifx\csname bibfnamefont\endcsname\relax
  \def\bibfnamefont#1{#1}\fi
\expandafter\ifx\csname citenamefont\endcsname\relax
  \def\citenamefont#1{#1}\fi
\expandafter\ifx\csname url\endcsname\relax
  \def\url#1{\texttt{#1}}\fi
\expandafter\ifx\csname urlprefix\endcsname\relax\def\urlprefix{URL }\fi
\providecommand{\bibinfo}[2]{#2}
\providecommand{\eprint}[2][]{\url{#2}}

\bibitem[{\citenamefont{Pan et~al.}(2000)\citenamefont{Pan, Bouwmeester,
  Daniell, Weinfurter, and Zeilinger}}]{pan00}
\bibinfo{author}{\bibfnamefont{J.}~\bibnamefont{Pan}},
  \bibinfo{author}{\bibfnamefont{D.}~\bibnamefont{Bouwmeester}},
  \bibinfo{author}{\bibfnamefont{M.}~\bibnamefont{Daniell}},
  \bibinfo{author}{\bibfnamefont{H.}~\bibnamefont{Weinfurter}},
  \bibnamefont{and}
  \bibinfo{author}{\bibfnamefont{A.}~\bibnamefont{Zeilinger}},
  \bibinfo{journal}{Nature} \textbf{\bibinfo{volume}{403}},
  \bibinfo{pages}{515} (\bibinfo{year}{2000}).

\bibitem[{\citenamefont{Bourennane et~al.}(2004)\citenamefont{Bourennane, Eibl,
  Kurtsiefer, Gaertner, Weinfurter, G\"uhne, Hyllus, Bru\ss{}, Lewenstein, and
  Sanpera}}]{bourennane04}
\bibinfo{author}{\bibfnamefont{M.}~\bibnamefont{Bourennane}},
  \bibinfo{author}{\bibfnamefont{M.}~\bibnamefont{Eibl}},
  \bibinfo{author}{\bibfnamefont{C.}~\bibnamefont{Kurtsiefer}},
  \bibinfo{author}{\bibfnamefont{S.}~\bibnamefont{Gaertner}},
  \bibinfo{author}{\bibfnamefont{H.}~\bibnamefont{Weinfurter}},
  \bibinfo{author}{\bibfnamefont{O.}~\bibnamefont{G\"uhne}},
  \bibinfo{author}{\bibfnamefont{P.}~\bibnamefont{Hyllus}},
  \bibinfo{author}{\bibfnamefont{D.}~\bibnamefont{Bru\ss{}}},
  \bibinfo{author}{\bibfnamefont{M.}~\bibnamefont{Lewenstein}},
  \bibnamefont{and} \bibinfo{author}{\bibfnamefont{A.}~\bibnamefont{Sanpera}},
  \bibinfo{journal}{Phys. Rev. Lett.} \textbf{\bibinfo{volume}{92}},
  \bibinfo{pages}{087902} (\bibinfo{year}{2004}).

\bibitem[{\citenamefont{{Walther} et~al.}(2005)\citenamefont{{Walther},
  {Resch}, {Rudolph}, {Schenck}, {Weinfurter}, {Vedral}, {Aspelmeyer}, and
  {Zeilinger}}}]{walther05}
\bibinfo{author}{\bibfnamefont{P.}~\bibnamefont{{Walther}}},
  \bibinfo{author}{\bibfnamefont{K.~J.} \bibnamefont{{Resch}}},
  \bibinfo{author}{\bibfnamefont{T.}~\bibnamefont{{Rudolph}}},
  \bibinfo{author}{\bibfnamefont{E.}~\bibnamefont{{Schenck}}},
  \bibinfo{author}{\bibfnamefont{H.}~\bibnamefont{{Weinfurter}}},
  \bibinfo{author}{\bibfnamefont{V.}~\bibnamefont{{Vedral}}},
  \bibinfo{author}{\bibfnamefont{M.}~\bibnamefont{{Aspelmeyer}}},
  \bibnamefont{and}
  \bibinfo{author}{\bibfnamefont{A.}~\bibnamefont{{Zeilinger}}},
  \bibinfo{journal}{Nature} \textbf{\bibinfo{volume}{434}},
  \bibinfo{pages}{169} (\bibinfo{year}{2005}).

\bibitem[{\citenamefont{{Lu} et~al.}(2007)\citenamefont{{Lu}, {Zhou},
  {G{\"u}hne}, {Gao}, {Zhang}, {Yuan}, {Goebel}, {Yang}, and {Pan}}}]{lu07}
\bibinfo{author}{\bibfnamefont{C.}~\bibnamefont{{Lu}}},
  \bibinfo{author}{\bibfnamefont{X.}~\bibnamefont{{Zhou}}},
  \bibinfo{author}{\bibfnamefont{O.}~\bibnamefont{{G{\"u}hne}}},
  \bibinfo{author}{\bibfnamefont{W.}~\bibnamefont{{Gao}}},
  \bibinfo{author}{\bibfnamefont{J.}~\bibnamefont{{Zhang}}},
  \bibinfo{author}{\bibfnamefont{Z.}~\bibnamefont{{Yuan}}},
  \bibinfo{author}{\bibfnamefont{A.}~\bibnamefont{{Goebel}}},
  \bibinfo{author}{\bibfnamefont{T.}~\bibnamefont{{Yang}}}, \bibnamefont{and}
  \bibinfo{author}{\bibfnamefont{J.}~\bibnamefont{{Pan}}},
  \bibinfo{journal}{Nature Phys.} \textbf{\bibinfo{volume}{3}},
  \bibinfo{pages}{91} (\bibinfo{year}{2007}).

\bibitem[{\citenamefont{Widera et~al.}(2004)\citenamefont{Widera, Mandel,
  Greiner, Kreim, H\"ansch, and Bloch}}]{widera04}
\bibinfo{author}{\bibfnamefont{A.}~\bibnamefont{Widera}},
  \bibinfo{author}{\bibfnamefont{O.}~\bibnamefont{Mandel}},
  \bibinfo{author}{\bibfnamefont{M.}~\bibnamefont{Greiner}},
  \bibinfo{author}{\bibfnamefont{S.}~\bibnamefont{Kreim}},
  \bibinfo{author}{\bibfnamefont{T.~W.} \bibnamefont{H\"ansch}},
  \bibnamefont{and} \bibinfo{author}{\bibfnamefont{I.}~\bibnamefont{Bloch}},
  \bibinfo{journal}{Phys. Rev. Lett.} \textbf{\bibinfo{volume}{92}},
  \bibinfo{pages}{160406} (\bibinfo{year}{2004}).

\bibitem[{\citenamefont{{Anderlini} et~al.}(2007)\citenamefont{{Anderlini},
  {Lee}, {Brown}, {Sebby-Strabley}, {Phillips}, and {Porto}}}]{anderlini07}
\bibinfo{author}{\bibfnamefont{M.}~\bibnamefont{{Anderlini}}},
  \bibinfo{author}{\bibfnamefont{P.~J.} \bibnamefont{{Lee}}},
  \bibinfo{author}{\bibfnamefont{B.~L.} \bibnamefont{{Brown}}},
  \bibinfo{author}{\bibfnamefont{J.}~\bibnamefont{{Sebby-Strabley}}},
  \bibinfo{author}{\bibfnamefont{W.~D.} \bibnamefont{{Phillips}}},
  \bibnamefont{and} \bibinfo{author}{\bibfnamefont{J.~V.}
  \bibnamefont{{Porto}}}, \bibinfo{journal}{Nature}
  \textbf{\bibinfo{volume}{448}}, \bibinfo{pages}{452} (\bibinfo{year}{2007}).

\bibitem[{\citenamefont{{H{\"a}ffner} et~al.}(2005)\citenamefont{{H{\"a}ffner},
  {H{\"a}nsel}, {Roos}, {Benhelm}, {Chek-Al-Kar}, {Chwalla}, {K{\"o}rber},
  {Rapol}, {Riebe}, {Schmidt} et~al.}}]{haffner05}
\bibinfo{author}{\bibfnamefont{H.}~\bibnamefont{{H{\"a}ffner}}},
  \bibinfo{author}{\bibfnamefont{W.}~\bibnamefont{{H{\"a}nsel}}},
  \bibinfo{author}{\bibfnamefont{C.~F.} \bibnamefont{{Roos}}},
  \bibinfo{author}{\bibfnamefont{J.}~\bibnamefont{{Benhelm}}},
  \bibinfo{author}{\bibfnamefont{D.}~\bibnamefont{{Chek-Al-Kar}}},
  \bibinfo{author}{\bibfnamefont{M.}~\bibnamefont{{Chwalla}}},
  \bibinfo{author}{\bibfnamefont{T.}~\bibnamefont{{K{\"o}rber}}},
  \bibinfo{author}{\bibfnamefont{U.~D.} \bibnamefont{{Rapol}}},
  \bibinfo{author}{\bibfnamefont{M.}~\bibnamefont{{Riebe}}},
  \bibinfo{author}{\bibfnamefont{P.~O.} \bibnamefont{{Schmidt}}},
  \bibnamefont{et~al.}, \bibinfo{journal}{Nature}
  \textbf{\bibinfo{volume}{438}}, \bibinfo{pages}{643} (\bibinfo{year}{2005}).

\bibitem[{\citenamefont{{Leibfried} et~al.}(2005)\citenamefont{{Leibfried},
  {Knill}, {Seidelin}, {Britton}, {Blakestad}, {Chiaverini}, {Hume}, {Itano},
  {Jost}, {Langer} et~al.}}]{leibfried05}
\bibinfo{author}{\bibfnamefont{D.}~\bibnamefont{{Leibfried}}},
  \bibinfo{author}{\bibfnamefont{E.}~\bibnamefont{{Knill}}},
  \bibinfo{author}{\bibfnamefont{S.}~\bibnamefont{{Seidelin}}},
  \bibinfo{author}{\bibfnamefont{J.}~\bibnamefont{{Britton}}},
  \bibinfo{author}{\bibfnamefont{R.~B.} \bibnamefont{{Blakestad}}},
  \bibinfo{author}{\bibfnamefont{J.}~\bibnamefont{{Chiaverini}}},
  \bibinfo{author}{\bibfnamefont{D.~B.} \bibnamefont{{Hume}}},
  \bibinfo{author}{\bibfnamefont{W.~M.} \bibnamefont{{Itano}}},
  \bibinfo{author}{\bibfnamefont{J.~D.} \bibnamefont{{Jost}}},
  \bibinfo{author}{\bibfnamefont{C.}~\bibnamefont{{Langer}}},
  \bibnamefont{et~al.}, \bibinfo{journal}{Nature}
  \textbf{\bibinfo{volume}{438}}, \bibinfo{pages}{639} (\bibinfo{year}{2005}).

\bibitem[{\citenamefont{{Julsgaard} et~al.}(2001)\citenamefont{{Julsgaard},
  {Kozhekin}, and {Polzik}}}]{julsgaard01}
\bibinfo{author}{\bibfnamefont{B.}~\bibnamefont{{Julsgaard}}},
  \bibinfo{author}{\bibfnamefont{A.}~\bibnamefont{{Kozhekin}}},
  \bibnamefont{and} \bibinfo{author}{\bibfnamefont{E.~S.}
  \bibnamefont{{Polzik}}}, \bibinfo{journal}{Nature}
  \textbf{\bibinfo{volume}{413}}, \bibinfo{pages}{400} (\bibinfo{year}{2001}).

\bibitem[{\citenamefont{Horodecki et~al.}(1996)\citenamefont{Horodecki,
  Horodecki, and Horodecki}}]{Horodecki1996}
\bibinfo{author}{\bibfnamefont{M.}~\bibnamefont{Horodecki}},
  \bibinfo{author}{\bibfnamefont{P.}~\bibnamefont{Horodecki}},
  \bibnamefont{and}
  \bibinfo{author}{\bibfnamefont{R.}~\bibnamefont{Horodecki}},
  \bibinfo{journal}{Phys. Lett. A} \textbf{\bibinfo{volume}{223}},
  \bibinfo{pages}{1} (\bibinfo{year}{1996}).

\bibitem[{\citenamefont{Terhal}(2000)}]{Terhal2000}
\bibinfo{author}{\bibfnamefont{B.}~\bibnamefont{Terhal}},
  \bibinfo{journal}{Phys. Lett. A} \textbf{\bibinfo{volume}{271}},
  \bibinfo{pages}{319} (\bibinfo{year}{2000}).

\bibitem[{\citenamefont{Lewenstein et~al.}(2000)\citenamefont{Lewenstein,
  Kraus, Cirac, and Horodecki}}]{Lewenstein2000}
\bibinfo{author}{\bibfnamefont{M.}~\bibnamefont{Lewenstein}},
  \bibinfo{author}{\bibfnamefont{B.}~\bibnamefont{Kraus}},
  \bibinfo{author}{\bibfnamefont{J.}~\bibnamefont{Cirac}}, \bibnamefont{and}
  \bibinfo{author}{\bibfnamefont{P.}~\bibnamefont{Horodecki}},
  \bibinfo{journal}{Phys. Rev. A} \textbf{\bibinfo{volume}{62}},
  \bibinfo{pages}{52310} (\bibinfo{year}{2000}).

\bibitem[{\citenamefont{Cirac et~al.}(2002)\citenamefont{Cirac, Horodecki,
  Hulpke, Kraus, Lewenstein, Sanpera, and Bru{\ss}}}]{Cirac2002}
\bibinfo{author}{\bibfnamefont{J.}~\bibnamefont{Cirac}},
  \bibinfo{author}{\bibfnamefont{P.}~\bibnamefont{Horodecki}},
  \bibinfo{author}{\bibfnamefont{F.}~\bibnamefont{Hulpke}},
  \bibinfo{author}{\bibfnamefont{B.}~\bibnamefont{Kraus}},
  \bibinfo{author}{\bibfnamefont{M.}~\bibnamefont{Lewenstein}},
  \bibinfo{author}{\bibfnamefont{A.}~\bibnamefont{Sanpera}}, \bibnamefont{and}
  \bibinfo{author}{\bibfnamefont{D.}~\bibnamefont{Bru{\ss}}},
  \bibinfo{journal}{J. Mod. Opt.} \textbf{\bibinfo{volume}{49}},
  \bibinfo{pages}{1399} (\bibinfo{year}{2002}).

\bibitem[{\citenamefont{G\"uhne et~al.}(2002)\citenamefont{G\"uhne, Hyllus,
  Bru{\ss}, Ekert, Lewenstein, Macchiavello, and Sanpera}}]{Guhne2002}
\bibinfo{author}{\bibfnamefont{O.}~\bibnamefont{G\"uhne}},
  \bibinfo{author}{\bibfnamefont{P.}~\bibnamefont{Hyllus}},
  \bibinfo{author}{\bibfnamefont{D.}~\bibnamefont{Bru{\ss}}},
  \bibinfo{author}{\bibfnamefont{A.}~\bibnamefont{Ekert}},
  \bibinfo{author}{\bibfnamefont{M.}~\bibnamefont{Lewenstein}},
  \bibinfo{author}{\bibfnamefont{C.}~\bibnamefont{Macchiavello}},
  \bibnamefont{and} \bibinfo{author}{\bibfnamefont{A.}~\bibnamefont{Sanpera}},
  \bibinfo{journal}{Phys. Rev. A} \textbf{\bibinfo{volume}{66}},
  \bibinfo{pages}{62305} (\bibinfo{year}{2002}), ISSN
  \bibinfo{issn}{1094-1622}.

\bibitem[{\citenamefont{{Bakr} et~al.}(2009{\natexlab{a}})\citenamefont{{Bakr},
  {Gillen}, {Peng}, {F{\"o}lling}, and {Greiner}}}]{bakr09}
\bibinfo{author}{\bibfnamefont{W.~S.} \bibnamefont{{Bakr}}},
  \bibinfo{author}{\bibfnamefont{J.~I.} \bibnamefont{{Gillen}}},
  \bibinfo{author}{\bibfnamefont{A.}~\bibnamefont{{Peng}}},
  \bibinfo{author}{\bibfnamefont{S.}~\bibnamefont{{F{\"o}lling}}},
  \bibnamefont{and}
  \bibinfo{author}{\bibfnamefont{M.}~\bibnamefont{{Greiner}}},
  \bibinfo{journal}{Nature} \textbf{\bibinfo{volume}{462}}, \bibinfo{pages}{74}
  (\bibinfo{year}{2009}{\natexlab{a}}).

\bibitem[{\citenamefont{Soderberg et~al.}(2009)\citenamefont{Soderberg,
  Gemelke, and Chin}}]{soderberg09}
\bibinfo{author}{\bibfnamefont{K.-A.~B.} \bibnamefont{Soderberg}},
  \bibinfo{author}{\bibfnamefont{N.}~\bibnamefont{Gemelke}}, \bibnamefont{and}
  \bibinfo{author}{\bibfnamefont{C.}~\bibnamefont{Chin}}, \bibinfo{journal}{New
  J. Phys.} \textbf{\bibinfo{volume}{11}}, \bibinfo{pages}{055022}
  (\bibinfo{year}{2009}).

\bibitem[{\citenamefont{Briegel and Raussendorf}(2001)}]{Briegel01}
\bibinfo{author}{\bibfnamefont{H.~J.} \bibnamefont{Briegel}} \bibnamefont{and}
  \bibinfo{author}{\bibfnamefont{R.}~\bibnamefont{Raussendorf}},
  \bibinfo{journal}{Phys. Rev. Lett.} \textbf{\bibinfo{volume}{86}},
  \bibinfo{pages}{910} (\bibinfo{year}{2001}).

\bibitem[{\citenamefont{{Mandel} et~al.}(2003)\citenamefont{{Mandel},
  {Greiner}, {Widera}, {Rom}, {H{\"a}nsch}, and {Bloch}}}]{mandel03}
\bibinfo{author}{\bibfnamefont{O.}~\bibnamefont{{Mandel}}},
  \bibinfo{author}{\bibfnamefont{M.}~\bibnamefont{{Greiner}}},
  \bibinfo{author}{\bibfnamefont{A.}~\bibnamefont{{Widera}}},
  \bibinfo{author}{\bibfnamefont{T.}~\bibnamefont{{Rom}}},
  \bibinfo{author}{\bibfnamefont{T.~W.} \bibnamefont{{H{\"a}nsch}}},
  \bibnamefont{and} \bibinfo{author}{\bibfnamefont{I.}~\bibnamefont{{Bloch}}},
  \bibinfo{journal}{Nature} \textbf{\bibinfo{volume}{425}},
  \bibinfo{pages}{937} (\bibinfo{year}{2003}).

\bibitem[{\citenamefont{Jaksch et~al.}(1999)\citenamefont{Jaksch, Briegel,
  Cirac, Gardiner, and Zoller}}]{Jaksch99}
\bibinfo{author}{\bibfnamefont{D.}~\bibnamefont{Jaksch}},
  \bibinfo{author}{\bibfnamefont{H.-J.} \bibnamefont{Briegel}},
  \bibinfo{author}{\bibfnamefont{J.~I.} \bibnamefont{Cirac}},
  \bibinfo{author}{\bibfnamefont{C.~W.} \bibnamefont{Gardiner}},
  \bibnamefont{and} \bibinfo{author}{\bibfnamefont{P.}~\bibnamefont{Zoller}},
  \bibinfo{journal}{Phys. Rev. Lett.} \textbf{\bibinfo{volume}{82}},
  \bibinfo{pages}{1975} (\bibinfo{year}{1999}).

\bibitem[{\citenamefont{{Klinger} et~al.}(2010)\citenamefont{{Klinger},
  {Degenkolb}, {Gemelke}, {Brickman Soderberg}, and {Chin}}}]{klinger10}
\bibinfo{author}{\bibfnamefont{A.}~\bibnamefont{{Klinger}}},
  \bibinfo{author}{\bibfnamefont{S.}~\bibnamefont{{Degenkolb}}},
  \bibinfo{author}{\bibfnamefont{N.}~\bibnamefont{{Gemelke}}},
  \bibinfo{author}{\bibfnamefont{K.}~\bibnamefont{{Brickman Soderberg}}},
  \bibnamefont{and} \bibinfo{author}{\bibfnamefont{C.}~\bibnamefont{{Chin}}},
  \bibinfo{journal}{Rev. Sci. Instrum.} \textbf{\bibinfo{volume}{81}},
  \bibinfo{pages}{013109} (\bibinfo{year}{2010}), \eprint{0909.2475}.

\bibitem[{\citenamefont{{Sackett} et~al.}(2000)\citenamefont{{Sackett},
  {Kielpinski}, {King}, {Langer}, {Meyer}, {Myatt}, {Rowe}, {Turchette},
  {Itano}, {Wineland} et~al.}}]{Sackett2000}
\bibinfo{author}{\bibfnamefont{C.~A.} \bibnamefont{{Sackett}}},
  \bibinfo{author}{\bibfnamefont{D.}~\bibnamefont{{Kielpinski}}},
  \bibinfo{author}{\bibfnamefont{B.~E.} \bibnamefont{{King}}},
  \bibinfo{author}{\bibfnamefont{C.}~\bibnamefont{{Langer}}},
  \bibinfo{author}{\bibfnamefont{V.}~\bibnamefont{{Meyer}}},
  \bibinfo{author}{\bibfnamefont{C.~J.} \bibnamefont{{Myatt}}},
  \bibinfo{author}{\bibfnamefont{M.}~\bibnamefont{{Rowe}}},
  \bibinfo{author}{\bibfnamefont{Q.~A.} \bibnamefont{{Turchette}}},
  \bibinfo{author}{\bibfnamefont{W.~M.} \bibnamefont{{Itano}}},
  \bibinfo{author}{\bibfnamefont{D.~J.} \bibnamefont{{Wineland}}},
  \bibnamefont{et~al.}, \bibinfo{journal}{\nat} \textbf{\bibinfo{volume}{404}},
  \bibinfo{pages}{256} (\bibinfo{year}{2000}).

\bibitem[{\citenamefont{{Ac{\'{\i}}n} et~al.}(2001)\citenamefont{{Ac{\'{\i}}n},
  {Bru{\ss}}, {Lewenstein}, and {Sanpera}}}]{Acin2001}
\bibinfo{author}{\bibfnamefont{A.}~\bibnamefont{{Ac{\'{\i}}n}}},
  \bibinfo{author}{\bibfnamefont{D.}~\bibnamefont{{Bru{\ss}}}},
  \bibinfo{author}{\bibfnamefont{M.}~\bibnamefont{{Lewenstein}}},
  \bibnamefont{and}
  \bibinfo{author}{\bibfnamefont{A.}~\bibnamefont{{Sanpera}}},
  \bibinfo{journal}{Phys. Rev. Lett.} \textbf{\bibinfo{volume}{87}},
  \bibinfo{pages}{040401} (\bibinfo{year}{2001}),
  \eprint{arXiv:quant-ph/0103025}.

\bibitem[{\citenamefont{T\'oth and G\"uhne}(2005{\natexlab{a}})}]{tothprl05}
\bibinfo{author}{\bibfnamefont{G.}~\bibnamefont{T\'oth}} \bibnamefont{and}
  \bibinfo{author}{\bibfnamefont{O.}~\bibnamefont{G\"uhne}},
  \bibinfo{journal}{Phys. Rev. Lett.} \textbf{\bibinfo{volume}{94}},
  \bibinfo{pages}{060501} (\bibinfo{year}{2005}{\natexlab{a}}).

\bibitem[{\citenamefont{T\'oth and G\"uhne}(2005{\natexlab{b}})}]{toth05}
\bibinfo{author}{\bibfnamefont{G.}~\bibnamefont{T\'oth}} \bibnamefont{and}
  \bibinfo{author}{\bibfnamefont{O.}~\bibnamefont{G\"uhne}},
  \bibinfo{journal}{Phys. Rev. A} \textbf{\bibinfo{volume}{72}},
  \bibinfo{pages}{022340} (\bibinfo{year}{2005}{\natexlab{b}}).

\bibitem[{\citenamefont{Benenti et~al.}(2007)\citenamefont{Benenti, Csati, and
  Strini}}]{benenti07}
\bibinfo{author}{\bibfnamefont{G.}~\bibnamefont{Benenti}},
  \bibinfo{author}{\bibfnamefont{G.}~\bibnamefont{Csati}}, \bibnamefont{and}
  \bibinfo{author}{\bibfnamefont{G.}~\bibnamefont{Strini}},
  \emph{\bibinfo{title}{Principles of Quantum Computation and Information}},
  vol.~\bibinfo{volume}{II} (\bibinfo{publisher}{World Scientific},
  \bibinfo{address}{Singapore}, \bibinfo{year}{2007}).

\bibitem[{\citenamefont{{Bakr} et~al.}(2009{\natexlab{b}})\citenamefont{{Bakr},
  {Gillen}, {Peng}, {F{\"o}lling}, and {Greiner}}}]{greiner09}
\bibinfo{author}{\bibfnamefont{W.~S.} \bibnamefont{{Bakr}}},
  \bibinfo{author}{\bibfnamefont{J.~I.} \bibnamefont{{Gillen}}},
  \bibinfo{author}{\bibfnamefont{A.}~\bibnamefont{{Peng}}},
  \bibinfo{author}{\bibfnamefont{S.}~\bibnamefont{{F{\"o}lling}}},
  \bibnamefont{and}
  \bibinfo{author}{\bibfnamefont{M.}~\bibnamefont{{Greiner}}},
  \bibinfo{journal}{Nature} \textbf{\bibinfo{volume}{462}}, \bibinfo{pages}{74}
  (\bibinfo{year}{2009}{\natexlab{b}}).

\bibitem[{\citenamefont{Gottesman}(1998)}]{gottesman98}
\bibinfo{author}{\bibfnamefont{D.}~\bibnamefont{Gottesman}},
  \bibinfo{journal}{Phys. Rev. A} \textbf{\bibinfo{volume}{57}},
  \bibinfo{pages}{127} (\bibinfo{year}{1998}).

\end{thebibliography}

\end{document}